\documentstyle[eqsecnum,aps]{revtex}
\begin{document}
\title{Nonlinear theory of autooscillations of quasiplanar interface during
directional solidification}
\author{I.A.Lubashevsky}
\address{Laboratory of Synergetics, the Moscow State University,\\ Vavilova str.
46/92, Moscow, 117333, Russia}
\author{V.V.Gafiychuk}
\address{Instute for Applied Problems of Mechanics and Mathematics,\\National
 Academy of Sciencies of Ukraine, 3b~Naukova~str., Lviv, 290601, Ukraine}
\author{M.G.Keijan}
\address{Institute of High Temperature, Academy of Sciencies of Russian Federation,\\ 13/19
Izhorskaia str., Moscow, 127412 Russia}
\date{\today}
\maketitle

\begin{abstract}
Within the framework of the frozen temperature approximation we develop a
strongly--nonlinear theory of one--dimensional pattern formation during
directional solidification of binary mixture under nonequilibrium
segregation. In the case of small partition coefficient the full problem is
reduced to the system of two ordinary differential equations describing the
interface motion in terms of its velocity and position coordinate. The type
of the oscillatory instability bifurcation is studied in detail in different
limits. For the subcrytical bifurcaton relaxation interface oscillations are
analyzed analytically and numerically. We show that these oscillation exibit
a number of anomalous properies. In particular, such oscillations can be
weakly-- or strongly--dissipative depending on the physical parameters and
the amplitude of the strongly--dissipative oscillations is determined not
only by the form of the corresponding nullcline but also by the behavior of
the system for small values of the interface velocity. Characteristic
parameters of the superlattice occuring in the growing crystal are estimated.
\end{abstract}



\section{INTRODUCTION}
\label{sec:int}

A great variety of patterns can form during directional solidification of
binary mixture when steady state motion of the planar solid--liquid
interface becomes unstable (for a review see, e.g., \cite{L80,WS91}). There
are several different mechanisms of this instability. One of them, proposed
by Mullins and Sekerka \cite{MS64} and called the marginal (morphological)
instability, has been investigated and discussed intensively for last years
\cite{WS91}. By this mechanism, under appropriate conditions the planar
interface loses stability mainly with respect to essentially nonuniform
perturbations, which gives rise to cellular interface formation.

Coriell and Sekerka \cite{CS83} have shown that departure from local
equilibrium at the planar solid/liquid interface due to high solidification
rates can cause oscillatory instability of its steady state motion. Merchant
and Davis \cite{MD90} analyzed this instability in detail taking into
account different characteristics of non-equilibrium conditions in a
thermodynamically-consistent way. Such an oscillatory instability can
manifest itself in self-formation of structures parallel to the moving
interface, which have been observed experimentally for Ag--Cu, Al--Cu, and
Al--Fe alloys solidified at high rates \cite{BSSB84,ZCK89,GCK91} (for a
review of materials exhibiting formation of this type structures see \cite
{KT90}). It should be noted that there are also other models predicting
oscillatory instability of plane solidification front \cite
{F82,SW83,CCR83,CCR86,BM84,BM91,GT82,GT84,J90,YuK91,G91}.

When solidification proceeds by the normal growth mechanism, non-equilibrium
effects at the solid/liquid interface become remarkable only at extremely
high solidification rates ($\sim $cm/sec--m/sec) \cite{BP85}. In this case
within the framework of classic approach \cite{CS83,MD90} departure from
local equilibrium at the planar interface due to its rapid motion is
characterized by the interface temperature $T_i(v,C_i)$ depending on the
interface velocity $v$ as
\begin{equation}
\label{1.1}T_i=T_0+m(v)C_i-\frac v\eta
\end{equation}
and the partition coefficient $k(v)={C_s}/{C_i}$ is also treated as a
certain function of $v$. Hear $T_0$ is the melting temperature of the pure
material being in thermodynamical equilibrium, $C_s$ and $C_i$ are the
interfacial concentrations of solute in the solid and liquid, respectively, $%
\eta $ is the interface mobility, and $m(v)$ is the nonequilibrium slope of
the liquidus regarded as a function of the interface velocity $v$. Several
models for the $k(v)$ dependence at high solidification rates have been
proposed \cite{W80,JGL80,A82} and the expression
\begin{equation}
\label{1.2}k(v)=\frac{k_e+\beta _0v}{1+\beta _0v}
\end{equation}
is much used in theoretical analysis of pattern formation during rapid
directional solidification. Here $k_e$ is the equilibrium partition
coefficient usually much less than unity, $k_e\ll 1$, and $1/\beta _0$ is
the characteristic velocity being typically of the size 5 m/sec \cite{BP85}.
The functions $m(v)$ and $k(v)$ are not independent but related by the
following expression \cite{BP85,BC86}
\begin{equation}
\label{1.3}m=m_e\biggl\{ 1+\frac{k_e-k[1-\ln (k/k_e)]}{1-k_e}\biggr\} ,
\end{equation}
where $m_e$ is the equilibrium value of $m$.

In mathematical analysis of directional solidification it is conventionally
used the ``frozen temperature'' approximation. In this approximation the
latent heat diffusion is neglected and the thermal properties of the liquid
and solid are taken to be equal, which decouples the temperature and solute
fields and enables one to represent the temperature distribution $T(y)$ in
the frame moving at the pulling velocity $V$ in the direction $y$ in the
form
\begin{equation}
\label{1.4}T(y)=T_0+Gy,
\end{equation}
where $G$ is the temperature gradient at the interface and the origin of the
moving coordinate system is chosen so that $\left. T(y)\right| _{y=0}=T_0$.
Sivashinsky and Novick-Cohen \cite{N-CS86,N-C87} have shown that the
difference in magnitude of the temperature diffusivity in solid and liquid
gives rise to an essentially nonlocal self--interaction of the interface in
addition to the ordinary renormalization of the temperature gradient $G$
\cite{MS64}. The effect of latent heat diffusion on rapid solidification has
been investigated in detail by Huntley and Davis \cite{HD?}, and Karma and
Sarkissian \cite{KS93}. It has been found that latent heat diffusion can
suppress the longwave oscillatory instability at high solidification rates.
However, because of the solute diffusivity $D$ in liquid being much less
than the temperature diffusivity $D_{\text{T}}$ ( $D\ll $ $D_{\text{T}}$)
the neighborhood of the zero wave number, where the latent heat effect is
considerable, is extremely narrow. The latter enables the temperature $T_i$
at the interface to be related to its coordinate $\zeta (t)$ along the $y$%
--axis of the moving frame by the expression \cite{KS93}
\begin{equation}
\label{1.5}T_i(t)=T_0+G\zeta (t)+\frac L{c_{\text{p}}}\int_{-\infty }^t\frac{%
dt^{\prime }}{\sqrt{4\pi D_{\text{T}}(t-t^{\prime })}}\frac{d\left\langle
\zeta (t^{\prime })\right\rangle }{dt^{\prime }}.
\end{equation}
Here $c_{\text{p}}$ is the specific heat, $L$ is the latent heat of fusion,
and $\left\langle \zeta (t)\right\rangle $ is the averaged position of the
interface at time $t$. When the mean curvature $K$ of the interface is not
extremely small so that $Kh_{\text{T}}\gg 1$ but $Kh\ll 1$, where $h_{\text{T%
}}\sim D_{\text{T}}/V$ and $h\sim D/V$ are the characteristic lengths of
heat and solute diffusion, respectively, the integral term in expression (%
\ref{1.5}) can be ignorable. In fact, in this case different interface
segments of size $1/K$ may move independently of one another, thus, their
total contribution to the integral term should be small, and the latent heat
diffusion controls solely their averaged position $\left\langle \zeta
(t)\right\rangle $ along the $y$--axis, fixing the value $\left\langle \zeta
(t)\right\rangle $. Under these conditions, as follows from (\ref{1.5}), the
``frozen temperature'' approximation holds also for rapid solidification, at
least, in semiquantitative analysis. The fact that longitudinal structures
forming during rapid solidification are quasi-plane rather than rigorously
plane is justified not only by the available experimental data but also
numerically\cite{KU86}.

When crystals grow layer--by--layer, with their surface being practically
parallel to one of the singular faces, the solute partitioning can become
non-equilibrium at slow rates typically used in technological processes,
viz., at $V\sim 10^{-3}$--$10^{-2}$ cm/sec \cite{V75}. The matter is that
the crystal growth at the volumetric solidification rate $V$ is attained
when atomic steps move along the crystal surface at the mean speed $V_{\text{%
step}}\sim V/\theta _{\text{cr}}$, where $\theta _{\text{cr}}$ is the mean
angle of interface misorientation from the ideal singular face. For real
singular faces forming in crystallization of monocrystals $\theta _{\text{cr}%
}\sim 10^{-3}-10^{-4}$ \cite{V75}, thus, $V_{\text{step}}\sim 1-100$ cm/sec
for these values of $V$. In the layer--by--layer crystallization solute
segregation to the solid involves two stages: surface trapping of solute
atoms by the moving steps and their following incorporation into the crystal
bulk. For such large values of $V_{\text{step}}$ the former process becomes
essentially non-equilibrium, causing the total partition coefficient $k(v)$
to depend on the interface velocity $v$. As shown by Voronkov \cite{V75} in
this case the $k(v)$ dependence can be represented in terms of
\begin{equation}
\label{1.6}k(v)=\frac{k_e+k_s\beta _s(v)}{1+\beta _s(v)},
\end{equation}
where the function $\beta _s(v)$ is of the form
\begin{equation}
\label{1.7}\beta _s(v)=\beta _1v\left[ 1+\beta _2v+\frac{k_e\beta _2}{%
k_s\beta _1}\right] ,
\end{equation}
the constants $\beta _1$, $\beta _2$ are of order $10^{2}$ sec/cm, $k_e$, as
before, is the equilibrium partition coefficient, $k_s$ is the interface
partition coefficient, and typically $k_e\ll k_s<1$. The given velocity
dependence of the partition coefficient for Al and P in Si has been observed
by Voronkov {\it et al}. \cite{VGSh67}. When the growing interface deviates
from the singular face the value $\theta _{\text{cr}}$ is ordinary about 1$^{%
\text{o}}$ or large and the solute partitioning can become non-equilibrium
at substantially higher rates of crystallization. A singular crystal surface
corresponds to a cusped minimum of the surface energy and, thereby, to an
extremely large value of the surface tension \cite{Ch80}. The latter
suppresses the marginal instability and practically does not affect the
oscillatory instability because at such low crystallization rates the effect
of latent heat diffusion should not be strong enough. In this case the
non--equilibrium solute partitioning through the induced oscillatory
instability can manifest itself in formation of solute bands, where the
solute concentration varies only in the growth direction perpendicular to
the interface, with the solute bands occurring near singular faces only.
Such pattern formation has been found to form during crystallization of InSb
with impurities Se and Te in addition to ordinary solute layers caused by
temperature fluctuations due to convective instabilities (see \cite{WG66,M91}
and references therein).

There are a number of works concerned with a nonlinear theory of pattern
formation in rapid solidification. Braun and Davis \cite{BD91,BD92} and
Braun {\it et al.} \cite{BMD92} developed a weakly-nonlinear theory and
derived the corresponding Ginzburg--Landau equation in the ``frozen
temperature'' approximation. Huntley and Davis \cite{HD93} incorporated in
this theory latent heat diffusion. Novick-Cohen \cite{N-C87} obtained a
Kuramoto--Sivashinsky equation in the limit $k\rightarrow 1$. Based on the
theory of oscillations with weak damping \cite{BH88} Merchant {\it et al.}
\cite{MBBD92} have studied nonlinear behavior of the zero-wavenumber
oscillatory instability in a small neighborhood of the absolute stability
point where dissipation processes are not pronounced. The same problem was
analyzed numerically by Brattkus and Meiron \cite{BM92}. Although the found
phase paths in the $v\zeta $--plane corresponding to the interface
oscillations deviate from the ellipsoidal shape substantially their theory
describes actually small amplitude interface oscillations as witnessed by
the existence of unclosed phase paths in the $v\zeta $--plane \cite{MBBD92}.
The appearance of unclosed phase paths in a small neighborhood of the
threshold and the transition from small amplitude (but may be
strongly--nonlinear) oscillations to really relaxation oscillations have
been analyzed in detail by Baer and Erneux \cite{BE86,BE92} having
considered as an example the well known ``French duck solution'' of the
FitzHugh--Nagumo equation. In \cite{GLO92} the authors of the present paper
actually following the boundary layer model proposed by Langer {\it et al.}
\cite{B-JGLS84} for directional solidification have studied strongly
nonlinear large amplitude oscillations of planar interface and numerically
found anomalous behavior of the corresponding limit cycle. It turns out that
although the corresponding nullcline is of the ``N--form'' the limit cycle
goes near only one of its stable branches whereas in the region containing
the other stable branch the phase path deviates from it substantially. It
should be noted that in the phenomenological model for banded structure
formation by Carrad {\it et al.} \cite{CGZK92} the phase path has been
assumed to follow the nullcline in the conventional way.

In the present paper we develop a theory of strongly nonlinear relaxation
oscillations of quasiplanar solid/liquid interface which can occur during
directional solidification. The term ``quasiplanar'' implies that the
interface can be treated as planar on spatial scales of order $h\sim D/V$ .
Since $D_{\text{T}}\gg D$ and so $h_{\text{T}}\gg h$ such an interface does
not have to be regarded as plane from the heat transfer standpoint, thus, we
may assume that the characteristic curvature radius $R$ of this interface
meets the inequality $h\ll R\ll h_{\text{T}}$. Besides, as has been
mentioned above real longitudinal structures forming during directional
solidification are not rigorously plane as well as those observed in
numerical experiments \cite{KU86}. This enables us to make use of the
``frozen temperature'' approximation, namely, we suppose that the interface
temperature $T_i$ and the interface position $\zeta $ in the coordinate
system moving at the pulling velocity $V$ are related by expression (\ref
{1.4}). Dealing with rapid directional solidification this approximation can
be justified, at least, in semiquantitive analysis especially if we consider
such values of the basic physical parameters that do not correspond to a
neighborhood of the nose of the neutral stability curve, where latent heat
diffusion affects solidification process most strongly \cite{HD93,HD?}.
Besides, in this case to a first approximation in the small parameter $\zeta
/R$ the nonlocal self--interaction of the interface due to difference in the
temperature diffusivities of the solid and liquid vanishes \cite{N-CS86}.
When crystals grow layer--by--layer approximation (\ref{1.4}) is also
justified because nonequilibrium partitioning comes into play even at low
crystallization rates \cite{V75} at which the latent heat effect is not yet
pronounced.

The second assumption adopted in the present analysis is that the interface
remains quasi-planar during oscillations. This can be the case when, in
particular, the marginal instability is suppressed by large values of
surface tension such that the absolute stability limit for the marginal
instability lies above that of the oscillatory instability. Such conditions
and the corresponding physical parameters have been analyzed in \cite{MD90}
for rapid solidification. For the layer--by--layer growth the given
assumption is justified to a larger degree because a singular crystal
surface corresponds to a cusped minimum of the surface energy and, so, to
extremely great values of the surface tension \cite{Ch80}. The given
assumption allows us to confine ourselves to one--dimensional equations
governing solute distribution in the melt.

The third basic assumption to be used is the smallness of the solute
partition coefficient $k(v)$ for all values of the interface velocity $v$
attained during oscillations. For rapid solidification problem this means
that we shall consider only the limit of relatively low solidification
rates. Concerning with the layer--by--layer growth this assumption does not
impose such exacting requirements. Indeed, in this case, as follows from
formula (\ref{1.6}), $k(v)\ll 1$ for all values of $v$ if $k_e$, $k_s\ll 1$.
The latter inequality is justified, e.g., for such crystals as Si and Ge
\cite{V75}. Besides, in the present analysis we shall ignore the dependence
of the liquidus slope $m$ on the interface velocity $v$ because for small
values of the partition coefficient $k(v)$ its variations should not lead to
appreciable variations of $m$ as results from expression (\ref{1.3}). The
assumption of the partition coefficient $k(v)$ being small allows the solute
distribution in the melt to be treated as quasistationary and enables us to
reduce the original problem to a system of two ordinary differential
equations whose solution can be investigated by the standard methods of
nonlinear analysis. As will be seen from the results to be obtained below
the general properties of interface oscillations are not sensitive to a
particular form of the $k(v)$ dependence. Keeping the latter in mind, in the
following analysis for the sake of definiteness we shall use such a function
$k(v)$ that generalizes expressions~(\ref{1.2}) and (\ref{1.6}) in a simple
way. Besides, for crystals growing layer--by--layer and by the normal growth
mechanism the velocity dependence of the interface temperature $T_i(v)$ is
different. Nevertheless, for the same reason we shall use expression~(\ref
{1.1}).

The paper is organized as follows. In Sec.~\ref{sec:ge} we review the
equations specifying the directional solidification problem under
consideration, briefly discuss linear stability of the steady--state
interface motion and recall under what conditions the interface may be
treated as quasiplanar. In Sec.~\ref{sec:qa} within the framework of the
quasistationary approximation the full diffusion problem is reduced to the
system of two ordinary differential equations. This section also includes
the bifurcation analysis of interface oscillations in order to compare the
results obtained in the quasistationary approximation with the previous
ones. In Sec.~\ref{sec:ro} we investigate the characteristic properties of
the interface relaxation oscillations that occur providing the bifurcation
is subcritical and estimate the corresponding parameters of the longitudinal
structure forming during solidification.

\section{GOVERNING EQUATIONS. THE COEXISTENCE OF DIFFERENT TYPE INSTABILITIES
}
\label{sec:ge}

We confine ourselves to the classic model for directional solidification of
a binary mixture \cite{L80,WS91,CS83,MD90}. A more introductory and detailed
exposition of this process can be found in \cite{L80,WS91,MD90,N-CS86,KS93}
and here we only summarize the basic equations governing interface motion
and our notation.

The solute concentration $C$ in the melt obeys the equation:

\begin{equation}
\label{2.1}\frac{\partial C}{\partial t}=D\nabla ^2C,
\end{equation}
where $D$ is the chemical diffusivity in the melt. Far in front of the
solid/liquid interface ${\cal I}$ the concentration $C$ tends to $C_\infty $
\begin{equation}
\label{2.2}C\rightarrow C_\infty \qquad \text{as}\qquad y\rightarrow \infty .
\end{equation}
The solute diffusion in the solid is neglected. Local conservation of solute
at the interface ${\cal I}$ leads to the following boundary condition:
\begin{equation}
\label{2.3}v_nC_i[1-k(v_n)]=\left. -D\nabla _nC\right| _{{\cal I}}.
\end{equation}
Here $v_n$ is the normal interface velocity, $C_i$ is the solute
concentration at the interface ${\cal I}$ on the melt side and $k(v_n)$ is
the partition coefficient. The velocity dependence of partition coefficient $%
k(v_n)$ due to nonequilibrium solute segregation is specified by
the expression
\begin{equation}
\label{k}k(v)=\frac{k_e+k_s\beta ^{*}v}{1+\beta ^{*}v},
\end{equation}
where $\beta ^{*}$ is a given constant. This formula is practically just a
simple generalization of expressions (\ref{1.2}) and (\ref{1.6}).

Keeping in mind the aforesaid in Sec.~\ref{sec:int} we confine our
consideration to the ``frozen temperature'' approximation in which the
temperature $T_i$ at the interface ${\cal I}$ and its $y$--coordinate $\zeta
$ in the frame moving at the pulling velocity $V$ in the $y$--direction are
related by the expression
\begin{equation}
\label{2.5}T=T_0+G\zeta .
\end{equation}
Here $T_0$ is a reference temperature chosen to be equal to the
solidification temperature of pure melt ($C=0$) and $G$ is the imposed
temperature gradient.

Following \cite{WS91,CS83,MD90} we suppose that the deviation of the
interface temperature $T_i$ from the local equilibrium value $T_{\text{eq}}$
causes the crystal growth at the velocity:

\begin{equation}
\label{2.6}v_n=\eta (T_{\text{eq}}-T_i),
\end{equation}
where $\eta $ is the interface mobility assumed to be isotropic and
constant. The local equilibrium temperature $T_{eq}$, solute concentration $%
C_i$ at the interface ${\cal I}$ and the interface curvature $H$ are related
by the Gibbs-Thomson condition:

\begin{equation}
\label{2.7}T_{\text{eq}}=T_0(1-dH)-mC_i,
\end{equation}
where $m>0$ is the slope of the liquidus line in the binary phase diagram,
and $d$ is the capillary length which is proportional to the surface tension
and accounts for anisotropic effects. The value of $m$ may be treated as
constant for such $v_n$ that $k(v_n)\ll 1$. In paper \cite{MD90} the
interface mobility $\eta $ has been represented actually as $\eta =V_0/m$.

Expressions (\ref{2.1})--(\ref{2.7}) form the complete model for the
directional solidification process under consideration.

The steady-state solution of equations (\ref{2.1})--(\ref{2.7}) describes
the motion of the planar interface ${\cal I}$ at the constant velocity $v^{%
\text{st}}=V$, the uniform distribution of impurity in the solid $C_{\text{%
sol}}^{\text{st}}=C_\infty $, and the stationary solute distribution in the
liquid , $y\geq \zeta $%
\begin{equation}
\label{3.1}C^{\text{st}}=C_\infty \left( 1+(\frac 1{k(V)}-1)\exp [-(y-\zeta )%
\frac VD]\right) .
\end{equation}

Under certain conditions the steady--state motion of the interface loses
stability and two different type instabilities, oscillatory and marginal,
can occur \cite{MS64,CS83}. The coexistence of these instabilities has been
analyzed in general in \cite{CS83,MD90,KS93,HD93}. However for the purpose
of the present paper it will be the best to consider briefly this problem in
the limit of small partition coefficient $(k\ll 1)$.

As follows from the results to be obtained below the amplification rate $%
\sigma $ of a small perturbation $\xi (x,t)\sim \exp [\frac VD(\sigma
Vt+ipx)] $ of the planar interface (where $p$ is the dimensionless wave
number) can be considered to be substantially less than one, $|\sigma |$$\ll
1,$ as long as $k(v)\ll 1$. In this limit the standard linear stability
analysis leads to the following eigenvalue equation:
\begin{equation}
\label{3.2}\sigma ^2(1+\mu )+\sigma \,\Im (p)+\Re (p)=0,
\end{equation}
where the functions
\begin{eqnarray*}
\Im (p) & = & \frac 1M+\mu k-Vk_V^{\prime }\,+\,(2+\gamma +\mu )\,p^2, \\
\Re (p) & = & \frac kM\,+\,[\frac 1M+\gamma k-1]\,p^2\,+\,\gamma \,p^4,
\end{eqnarray*}
and the parameters%
$$
M=\frac{VmC_\infty }{GDk},\quad \gamma =\frac{T_0dVk}{DmC_\infty },\quad \mu
=\frac{Vk}{\eta mC_\infty }.
$$
It should be noted that Merchant and Davis \cite{MD90} used practically the
same parameters where, however, the partition coefficient takes the
equilibrium value $k_e$.

According to equation~(\ref{3.2}) the planar interface becomes unstable
(i.e. Re$\sigma >0$) with respect to the given perturbation when $\Re (p)<0$
or $\Im (p)<0$. The former inequality is associated with the marginal
instability, which is caused by the constitutional undercooling and has been
thoughtfully analyzed and discussed (for a review see \cite
{L80,WS91,CS83,PPK91}). The corresponding maximum of the amplification rate $%
\sigma _{\text{max}}$ is attained at $p=p_{\text{max}}\not =0$ and, thereby,
the neutral curve I, separating stable and unstable (from the marginal
instability standpoint) regions in the $C_\infty V$--plane, depends on the
capillary length $d$ (Fig.~\ref{F.3}).
This instability region is specified by the inequality
$\Re (p_{\text{max}})<0$, leading to the expression

\begin{equation}
\label{3.5}mC_\infty \,\geq \,k\biggl\{ \frac{DG}V\,+\,\frac{kT_0dV}D\,+\,2%
\sqrt{T_0Gdk}\biggr\}.
\end{equation}
The latter inequality is bound up with the oscillatory instability which is
described in \cite{CS83,MD90} and occurs through the dependence of the
partition coefficient $k(v)$ on the velocity $v$. Since the function $\Im
(p) $ is increasing the corresponding maximum of amplification rate $\sigma $
is attained at $p=0$, thus in the ``frozen temperature'' approximation the
oscillatory instability region in the $C_\infty V$ - plane is determined by
the inequality

\begin{equation}
\label{3.6}mC_\infty \,\geq \,\frac k{Vk_V^{\prime }}\left\{ \frac{DG}V\,+\,%
\frac{Vk}\eta \right\}
\end{equation}
and is independent of the capillary length $d$. The latent heat diffusion
can suppress the oscillatory instability for small wave numbers \cite
{KS93,HD93}. However due to the heat diffusivity being much larger than the
solute one the wave number interval where the latent heat essentially
affects perturbation dynamics is narrow enough. Therefore, the condition (%
\ref{3.6}) obtained in the ``frozen temperature'' model holds true if the
latent heat fusion is not extremely great. The equality in (\ref{3.6})
specifies the neutral curve II of the oscillatory instability
(Fig.~\ref{F.3}).

In particular, it follows from (\ref{3.6}) that for arbitrary values of the
pulling velocity $V$ and the imposed temperature gradient $G$ all
disturbances corresponding to the oscillatory instabiblity decay when $%
C_\infty <C^{*}$, where
\begin{equation}
\label{as}C^{*}=\frac{k_e^2}{m\eta (k_s-k_e)\beta ^{*}}.
\end{equation}
In other words, the inequality $C_\infty <C^{*}$ is the absolute stability
criterion for the oscillatory instability in the limit of small partitition
coefficients. It should be noted that this criterion matches the results
obtained by Merchant and Davis \cite{MD90} in the limit $V\rightarrow 0$.
For $C_\infty >C^{*}$ the oscillatory instability can occur only in the
region $C_\infty >C_{\text{as}}(\beta )$ specified by the function $C_{\text{%
as}}(\beta )$ of the dimensionless pulling velocity $\beta =\beta ^{*}V$%
\begin{equation}
\label{vc}C_{\text{as}}(\beta )=C^{*}(1+\frac \beta r)^2,
\end{equation}
where the parameter $r=k_e/k_s$. In the $C_\infty V$--plane the curve $%
C_\infty =C_{\text{as}}(\beta )$ is the boundary of the absolute stability
region (line~1 in Fig.~\ref{F.2}). In this figure line~2
represents the neutral curve of the oscillatory instability for a fixed
value of $G$ measured in units of $G^{*}=k_e/(D\eta \beta ^{*2})$. In
particular, in these terms the neutral curve can be represented as
\begin{equation}
\label{ncI}C_\infty =C_{\text{as}}(\beta )\left[ 1+\frac G{G^{*}}\frac{%
1+\beta }{\beta ^2(1+\beta /r)}\right] .
\end{equation}
The neutral curve in contrast to the absolute stability curve attains its
minimum at $V=V_{\text{ext}}>0$. For example, as follows from (\ref{ncI}) when
$r\ll 1$ and $r^2\ll G/G^{*}\ll 1/r$ the value $V_{\text{ext}}$ belongs to
the region $r^2/{\beta^{*}}\ll V\ll 1/{\beta^{*}}$ and so
\begin{equation}
\label{vmin}V_{\text{ext}}\approx \frac1{\beta^{*}} \biggl(\frac {rG}{2G^{*}}%
\biggr).
\end{equation}

As the capillary length increases the neutral curve I for a fixed value of $G
$ goes upwards in the $C_\infty V$--plane and, thus, there can be three
characteristic relative positions of curves I and II shown in Fig.~\ref{F.3}. When
the capillary length is small enough ($d\ll d_1$, where $d_1=d_1(D,G,\eta
,T_0)$ is such a value of the capillary length that curve I crosses curve II
approximately at its minimum point) the region of the oscillatory
instability is totally located inside that of the marginal instability. In
this case the oscillatory instability of the planar interface seems not to
be able to manifest itself in any way because of rapid development of
spatially nonuniform interface perturbations (Fig.~\ref{F.3}(a)). When $%
d\sim d_1$ there is a certain small region in the $C_\infty V$--plane that
corresponds solely to the oscillatory instability (Fig.~\ref{F.3}(b)). In
this case the oscillatory instability can occur. However, as it has been
shown numerically \cite{KU86}, the interface becomes substantially nonplanar
during the instability development. In this case the banded structures seem
to occur \cite{MD90}. For $d\gg d_1$ the large value of the capillary
length, i.e. of the surface tension, suppresses the marginal instability and
thereby the interface tends to be planar. It should be noted that in this
case (Fig.~\ref{F.3}(c)) the developed form of the solidification front will
be quasiplanar and solute band formation is expected \cite{MD90}.

\section{QUASISTATIONARY APPROXIMATION}
\label{sec:qa}

In this section we obtain nonlinear evolution equations for the planar
interfacial position $y=\zeta (t)$. In general, these equations must contain
memory effects; that is a displacement of the interface {$\cal I$} from its
steady-state position causes a perturbation of the impurity distribution
which, in turn, affects the motion of the interface at later times. However,
in the limit of small partition coefficient it turns out that the interface
velocity varies in time so slowly that it remains practically constant
during the time needed for relaxation of the solute distribution in the
melt. In other words, it means that all temporal scales of interface
dynamics are much larger than the characteristic time ($\tau _v\sim D/V^2$)
it takes for the steady state solute distribution to form when $C_i$ and $v$
are fixed.

In the frame of reference moving in the $y$--direction at the interface
velocity $v$ the diffusion equation has the form:

\begin{equation}
\label{4.1}\dot C=DC_{yy}+vC_y.
\end{equation}
Following \cite{B-JGLS84,KG85,WL93} we integrate (\ref{4.1}) over $y$ from $%
\zeta$ to $\infty $ then taking into account the boundary condition (\ref
{2.3}) we obtain

\begin{equation}
\label{4.2}\frac d{dt}\int _{\zeta}^\infty
dy[C(y,t)-C_\infty ]=v[C_\infty
-k(v)C_i].
\end{equation}
In the quasistationary approximation we can regard the left hand side of
equation~(\ref{4.1}) as a small perturbation and neglect it at lower order
of the perturbation technique. In other words, setting the transient term in
(\ref{4.1}) equal to zero and then solving the obtained equation we find the
lower approximation of $C(y,t)$: for $y>\zeta $

\begin{equation}
\label{4.3}C(y,t)=C_\infty +[C_i(t)-C_\infty ]\exp \left\{ -\frac{v(t)}D%
(y-\zeta )\right\}
\end{equation}
Substituting expression (\ref{4.3}) into equation~(\ref{4.2}) we get

\begin{equation}
\label{4.4}D\frac d{dt}\left[ \frac{C_i-C_\infty }v\right] =v[C_\infty
-k(v)C_i].
\end{equation}
Equation (\ref{4.4}) along with the equation

\begin{equation}
\label{4.5}\frac{dy}{dt}=v-V
\end{equation}
and the relationship

\begin{equation}
\label{4.6}v=-\eta (G\zeta +mC_i)
\end{equation}
resulting from (\ref{2.5})--(\ref{2.7}) form a complete description of the
quasiplanar interface motion.

The steady state solution of the given system of equations is of the form:

\begin{equation}
\label{4.6ad}v^{\text{st}}=V,\quad C_i^{\text{st}}=\frac{C_\infty }{k(V)}%
,\quad \zeta ^{\text{st}}=-\frac 1G\biggl[ \frac V\eta +\frac{mC_\infty }k%
\biggr].
\end{equation}
It is convenient for the most of the following analysis to use the
dimensionless variables, $a$ and $u$:

\begin{equation}
\label{4.7}a=\frac{\zeta-\zeta ^{\text{st}}}{\zeta^{\text{st}}}, \quad u=%
\frac{v-V}V
\end{equation}
In terms of this variables, the equations of motion become:

\begin{eqnarray}
\label{4.8}\tau \frac{du}{dt} & = & \frac{(1+u)^3}{(1+a)}\frac{k[u]}{k[0]}%
\left( a-a_0[u]\right),\\
\label{4.9}\tau \frac{da}{dt} & = & -(\tau \omega )^2u.
\end{eqnarray}
Here we have also introduced the quantities:

\begin{equation}
\label{4.9p}\tau =\frac D{V^2k},(\tau \omega )^2=\frac 1{kM(1+\mu )}%
,k[u]=k(V_0[1+u]),
\end{equation}
where $\omega $ is the frequency of the interface oscillations at the
threshold, and the function
\begin{equation}
\label{4.z}a_0[u]=\frac 1{(1+\mu )}\left( \frac{k[0]}{k[u]}+\mu u-1\right)
+(\tau \omega )^2\frac{k[0]}{k[u]}\frac u{(1+u)^2}
\end{equation}
specifies the nullcline of equation~(\ref{4.8}).

In this way the full diffusion problem of interface dynamics is reduced to
the system of ordinary differential equations, which describes the nonlinear
oscillations of the quasiplanar interface \cite{GLO92}.

In the present section on the basis of evolution equations (\ref{4.8}), (\ref
{4.9}) we also study the bifurcation mode of the oscillation onset,
depending on relation between physical parameters such as $V,G,\eta $ etc.
In addition, to validate the quasistationary approximation, we compare the
form of the interface auto-oscillations obtained in the quasistationary
approximation for the supercritical bifurcation and those obtained in the
one-dimensional full diffusion problem described in Sec.~\ref{sec:ge}.

In order to analyze the bifurcation mode we can make use of the
Bogolyubov--Krylov--Mitropolskii technique and the theory of Hopf
bifurcation \cite{BM74,HKW81}. In this way near the threshold the interface
motion is described by the quasiharmonic time dependence of the variable $%
a=A(t)\cos (\omega (t)t)$, where $\omega (t)\approx \omega $ and $A(t)$ are
certain functions of the time $t$ being practically constant on the temporal
scale $1/\omega $. Following the standard procedure \cite{BM74,HKW81} we
find that the amplitude $A(t)$ of the quasiharmonic oscillations obeys the
equations

\begin{equation}
\label{5.7}\tau \frac{dA}{dt}=\frac 12A(\epsilon +\alpha A^2),
\end{equation}
where the constants $\epsilon $ and $\alpha $ are specified by the formulae
\begin{equation}
\label{b1}\epsilon =-a_0^{\prime }[0]=\frac 1{(1+\mu )k[0]}\left( \mu k[0]+%
\frac 1M-k^{\prime }[0]\right) ,
\end{equation}
\begin{equation}
\label{b2}\alpha =-\frac 1{8(\tau \omega )^2}\left\{ a_0^{\prime \prime
\prime }[0]+2a_0^{\prime \prime }[0](3+\frac{k^{\prime }[0]}{k[0]})\right\} ,
\end{equation}
and the primes on the symbols denote the derivatives of the corresponding
functions with respect to the variable $u$ taken at the point $u=0$. In
these terms the instability condition of steady state interface motion takes
the form $a_0^{\prime }[0]<0$. In other words, for the oscillatory
instability to occur the nullcline $a=a_0[u]$ should be decreasing in the
vicinity of the steady state point ($u=0$, $a=0$). Taking into account (\ref
{4.z}) we can rewrite the latter inequality in the form $\mu
+1/(k[0]M)<k^{\prime }[0]/k[0]$ which as it must is exactly the
dimensionless form of inequality~(\ref{3.6}).

In the following analysis it will be convenient to measure the parameters $%
\mu $ and $1/(k[0]M)$ also in units of $k^{\prime }[0]/k[0]$. In other
words, let us introduce new parameters $\phi _g,\phi _\mu \geq 0$ defined by
the formulae
\begin{equation}
\label{5.102}\frac 1{k[0]M}=\phi _g\frac{k^{\prime }[0]}{k[0]},\quad \mu
=\phi _\mu \frac{k^{\prime }[0]}{k[0]}.
\end{equation}
In particular, at the points of the neutral curve $\phi _g+\phi _\mu =1$ and
at the absolute stability boundary, $\mu _{\text{as}}=k^{\prime }[0]/k[0]$,
we get $\phi _g=0$ and $\phi _\mu =1$. Then substituting (\ref{4.z}) into (%
\ref{b2}) and calculating the obtained result at the threshold, $a_0^{\prime
}[0]=0$, we find an expression for the parameter $\alpha $ which for the $%
k(v)$ dependence specified by formula (\ref{k}) takes the form
\begin{equation}
\label{5.101}\alpha =\alpha _0[u]\left( \phi -\phi _b[u]\right) .
\end{equation}
Here the parameter $\phi =\phi _g\in [0,1]$ characterizes deviation of the
system from the absolute stability boundary, $\alpha _0[u]$ is a certain
positively definite function of the variable $u$, and the function $\phi
_b[u]$ is given by the expression
\begin{eqnarray}
\label{5.103}
\lefteqn{\phi _b[\beta ]=\frac \beta {r+\beta }\biggl[ 6-\frac{2\beta }{1+\beta }-\frac \beta {r+\beta }\biggr]} \nonumber \\
& &{}\times \biggl[ 3+\frac{(1-r)\beta }{(1+\beta
)(r+\beta )}(4-\frac{2\beta }{1+\beta }-\frac \beta {r+\beta })\biggr]
^{-1}.
\end{eqnarray}
According to (\ref{5.103}) bifurcation of the oscillatory instability is
supercritical if $\phi <\phi _b[\beta ]$ and subcritical for $\phi >\phi
_b[\beta ]$. As follows from (\ref{5.103}), when $k_e\ll k_s$, i.e. $r\ll 1$
there are three limits characterizing different behavior of the function $%
\phi _b[\beta ]$. In the limit $\beta \ll r$ (i.e. $V\ll r/\beta ^{*}$) the
function $\phi _b[\beta ]\approx 2\beta /r$ which actually exactly matching
the results obtained by Huntley and Davis \cite{HD93} for small pulling
velocities in the frozen temperature approximation. When $r\ll \beta \ll 1$
(i.e. $r/\beta ^{*}\ll V\ll 1/\beta ^{*}$) the value $\phi _b[\beta ]\approx
5/6$ and for $\beta \sim 1$ ($V\sim 1/\beta ^{*}$) we get that $\phi
_b[\beta ]\rightarrow 1$ as $\beta \rightarrow 1$ and $\phi _b[\beta ]>1$
when $\beta >1$. In the latter case, $\beta >1$, bifurcation is always
supercritical. In the $C_\infty V$--plane the curve corresponding to the
transition point from supercritical to subcritical bifurcation is specified
by the formula $C_{\text{tr}}=C_{\text{as}}(\beta )/(1-\phi _b[\beta ])$
(for $\beta =\beta ^{*}V<1$) and is shown by the curve 3 in Fig.~\ref
{F.2}. For a fixed value of the temperature gradient $G$ the neutral curve
in the $C_\infty V$--plane crosses the curve $C_\infty =C_{\text{tr}}(\beta )
$ at a certain point $(V_c,C_c)$ and for $V<V_c$ as well as $V>V_c$ the
oscillatory instability onset at the threshold is characterized by
subcritical and supercritical bifurcation, respectively. In particular, when
$r^2\ll G/G^{*}\ll 1/r$ this point belongs to the region $r/\beta ^{*}\ll
V\ll 1/\beta ^{*}$ and in this case the values $V_c$, $V_{\text{ext}}$ are
related by the expression $V_c=(2/5)^{1/3}V_{\text{ext}}$.

In the region $V>V_c$ where bifurcation is supercritical and the coefficient
$\alpha $ is negative the amplitude of the steady-state quasiharmonic
oscillations $A_0$ near the threshold is

\begin{equation}
\label{5.14}A_0=\left( \frac \epsilon \alpha \right) ^{\frac 12}\propto
\left[ \frac{k^{\prime }[0]}{k[0]}-(\frac 1{k[0]M}+\mu )\right]^{1/2}.
\end{equation}
For $V<V_c$, the coefficient $\alpha $ is positive and in order to find the
amplitude of the interface oscillations a more complicated analysis is
required (this case is considered in Sec.~\ref{sec:ro}).

It should be noted that the results of the present section concerning the
bifurcation behavior are in qualitative agreement with those obtained in the
previous papers for different physical conditions \cite{J90,BD91,BD92,BMD92}.

To examine the feasibility of the quasistationary approximation in the
supercritical mode we compare the limit cycles of the interface
autooscillations as well as the time course $a(t)$ and $u(t)$ obtained by
solving equations (\ref{4.8}), (\ref{4.9}) and the full one-dimensional
problem (\ref{2.1})--(\ref{2.7}). For this purpose we have solved
numerically the equations mentioned above using the following values of the
parameters: $k_0=0.0125$,$\,k_\infty =0.8$,$\,\beta ^{*}V=0.125$,$\,kM=10$,
and $\mu =0.5$. In this case the ratio $k^{\prime }[0]/k[0]\simeq 0.75$ and,
thus, in the $C_\infty V$--plane the equilibrium point of the system is
practically located near the boundary of the oscillatory instability region
because of $[k^{\prime }/k-(k/M+\mu )]/(k^{\prime }/k)\sim 0.2$. The
obtained results are plotted in figures~\ref{F.4}--\ref{F.5} where the
dashed and solid lines corresponds to the solutions of equations (\ref{4.8}%
), (\ref{4.9}) and the full diffusion model, respectively.

As seen from Fig.~\ref{F.4} the quasistationary approximation gives an
adequate description of the interface oscillations at least in the
supercritical mode. Fig.~\ref{F.5} also demonstrates that even in the
vicinity of the instability region boundary the nonlinear effects play a
significant role in the interface oscillations and the limit cycle
substantially deviates from the elliptic shape and the time dependence $u(t)$%
, $a(t)$ are practically of the spikewise form. Nevertheless, the period $T$
of these interface autooscillations can be estimated by the expression $%
T=2\pi \omega ^{-1}=2\pi \tau [(1+\mu )kM]^{1/2}$ which is rigorously
justified for quasiharmonic oscillations only. Fig.~\ref{F.6} shows the
resulting impurity distribution in the solid $C_{\text{sol}}(y)$. According
to Fig.~\ref{F.6} and as would be expected the spatial period of the growing
superlattice is of order
\begin{equation}
\label{5.15}H\sim TV\sim 2\pi \Biggl[(1+\mu )\frac{DmC_\infty }{Vk^2G}\Biggr]%
^{1/2}.
\end{equation}

\section{RELAXATION OSCILLATIONS OF INTERFACE}
\label{sec:ro}

As shown in Sec.~\ref{sec:qa} the subcritical bifurcation corresponds to
small values of the pulling velocity, namely, $V<1/\beta ^{*}$ ($\beta <1$).
Besides, the asymptotic behavior of the curve separating in the $C_\infty V$%
--plane the regions of the subcritical and supercritical bifurcation is
different for $\beta \ll r$ and $r\ll \beta \ll 1$ (when $r=k_e/k_s\ll 1$).
So for $\beta <1$ the interface oscillations may be expected to be
relaxation, with their main features depending substantially on the ratio of
$\beta $ and $r$. In agreement with the results to be obtained below there
are two cases of strongly--nonlinear dynamics of the interface motion:
weakly-- and strongly--dissipative oscillations. Weakly--dissipative
oscillations (however of large amplitude with respect to $u$) can occur when
\begin{equation}
\label{6.100}\beta \ll r\quad \text{and}\quad \frac \beta r\ll (1-\phi _\mu
),\phi _g\ll \left( \frac \beta r\right) ^{1/3}.
\end{equation}
whereas the conditions
\begin{mathletters}
\begin{equation}
\label{6.100a}\beta \ll r\quad \text{and}\quad \left( \frac \beta r\right)
^{1/3}\ll (1-\phi _\mu ),\phi _g
\end{equation}
or
\begin{equation}
\label{6.100b}r\ll \beta \ll 1\quad (\text{for }r=k_e/k_s\ll 1)
\end{equation}
\end{mathletters}
will result in strongly--dissipative oscillations. It should be noted that
in the strict sense only strongly--dissipative oscillations are relaxation,
although in the two cases the velocity amplitude $u$ can attain large values
much greater than unity. In this section we analyze strongly--nonlinear
dynamics of the interface oscillations in the given cases individually.

Weakly--dissipative oscillations of crystal interface that can occur during
directional solidification has been analyzed in detail by Merchant et al.
\cite{MBBD92} and Brattkus and Meiron \cite{BM92} within the framework of
the classic model for rapid solidification similar to one used in the
present analysis. So we only briefly consider the first limit (\ref{6.100})
in order to complete the analysis of strongly--nonlinear dynamics of the
given model.

In this case the ratio
$$
\frac{k^{\prime }[0]}{k[0]}=\frac{(1-r)\beta }{(1+\beta )(r+\beta )}\approx
\frac \beta r\ll 1
$$
is a small parameter, $\mu \ll 1$, and the value
$$
(\tau \omega )^2\approx \phi _g\frac \beta r.
$$
According to Sec.~\ref{sec:qa}, for $\beta \ll r$ the bifurcation is
subcritical when $\phi _g>2\beta /r$ and at the threshold $\phi _g+\phi _\mu
=1$. So in order to analyze relaxation oscillations of the interface in the
given limit we may confine ourselves to the case $1-\phi _\mu \sim \phi _g$
and $\phi _g\gg \beta /r$. We note that the two relations have actually led
us to the left--hand side of the latter conditions of limit (\ref{6.100}).
Its right--hand side enables us to treat the ratio $a_0[u]/a$ in equation (%
\ref{4.8}) as a small parameter for the interface oscillations that will
occur under these conditions. The matter is that the characteristic value of
the amplitude $\bar a$ of the variable $a$ turns out to be about $\bar a\sim
\tau \omega $, so $\bar a\sim (\phi _g\beta /r)^{1/2}$ and the maximum $u_{%
\text{max}}$ of the interface velocity $u$ attained during oscillations can
be estimated as $u_{\text{max}}\sim \tilde u_{\text{min}}$, where
$\tilde u_{\text{%
min}}$ is the $u$--coordinate of the point $(\tilde u_{\text{min}},%
\tilde a_{\text{min}})$ at which the curve $a_0[u]$ attains its extremum
(minimum) in the region $u>0$ (Fig.~\ref{F.7}). In this case expression (\ref
{4.z}) may be rewritten in the form
\begin{equation}
\label{ncsu}a_0[u]\approx \frac \beta ru\left[ -(1-\phi _\mu )+\frac \beta r%
u+\phi _g\frac 1{(1+u)^2}\right] ,
\end{equation}
whence for $\phi _g\sim 1-\phi _\mu $ we get that
\begin{mathletters}
\label{adh.1}
\begin{equation}
\tilde u_{\text{min}} \approx \frac{r}{2\beta}(1-\phi_{\mu})\sim \frac{r}{%
\beta}\phi_g\gg 1,
\end{equation}
\begin{equation}
\tilde a_{\text{min}}
\approx  -\frac{1}{4}(1-\phi_{\mu})^2\sim
{\phi}^2_g,
\end{equation}
\end{mathletters}
so, by virtue of (\ref{6.100}), we obtain that $%
|a_0[u]|\lesssim |\tilde a_{\text{min}}|\ll \bar a$.

At lower order in the small parameter $a_0[u]/a$ the system of equations (%
\ref{4.8}), (\ref{4.9}) is conservative with the ``energy'' (the first
integral)
\begin{equation}
\label{6.h}{\cal H}(u,a)=\frac{(\tau \omega )^2}2\frac{u^2}{(1+u)^2}+\frac 12%
a^2
\end{equation}
When deriving expression (\ref{6.h}) we have also taken into account that
for the given values of the parameters $k[u]\approx k[0]$ and $a\ll 1$. The
remaining term in equation (\ref{4.8}) proportional to $a[u]$ causes time
variation of the ``energy'', namely,
\begin{equation}
\label{6.h1}\tau \frac{d{\cal H}}{dt}=-(\tau \omega )^2a_0[u]u.
\end{equation}

In the case under consideration the system dynamics may be described in
terms of fast motion along a phase path, specified by the equation ${\cal H}%
(u,a)=h$ for a fixed value of $h$, and slow time variations in the
``energy'' ${\cal H}(u,a)$. The geometry of the phase paths in the $ua$%
--plane is shown in Fig.~\ref{F.7}. As seen in Fig.~\ref{F.7} there are two
types of the phase paths, closed and unclosed, which correspond to $h<\frac 1%
2(\tau \omega )^2$ and $h>\frac 12(\tau \omega )^2$, respectively, and under
the adopted assumptions solely the closed paths are meaningful.

Steady--state oscillations matches such a value $h_{\text{st}}$ for which
the right--hand side of equation (\ref{6.h1}) averaged over a single period
of the system motion is equal to zero. Since the right--hand side of
equation (\ref{6.h1}) depends on the variable $u$ only, at lower
approximation we may average it assuming the system to move strictly along
the phase  for a certain fixed value of $h$. In this way from (\ref{6.h1})
we get
\begin{equation}
\label{6.h2}\tau \frac{dh}{dt}=\frac{2\tau }{T(h)}\int\limits_{u_{\text{min}%
}(h)}^{u_{\text{max}}(h)}du\frac{\partial a_{+}(h,u)}{\partial u}a_0[u],
\end{equation}
where $T(h)$ is the period of the system motion along the phase path
corresponding to ${\cal H}=h$, $u_{\text{min }}(h)$ and $u_{\text{max }}(h)$
are the minimum and maximum of the variable $u$ attained during the motion
along this phase path, and $a_{+}(h,u)$ is the positive solution of the
equation ${\cal H}(u,a)=h$, namely,
\begin{equation}
\label{6.h3}u_{\text{min}}(h)=-\frac{(2h)^{1/2}}{\tau \omega +(2h)^{1/2}}%
,\quad u_{\text{max}}(h)=\frac{(2h)^{1/2}}{\tau \omega -(2h)^{1/2}}
\end{equation}
and
\begin{equation}
\label{6.h4}a_{+}(h,u)=(\tau \omega )\left[ \frac{2h}{(\tau \omega )^2}-%
\biggl(\frac u{1+u}\biggr)^2\right] ^{1/2}.
\end{equation}
Substituting (\ref{6.h3}) and (\ref{6.h4}) into (\ref{6.h2}) and integrating
over $u$ we obtain
\begin{eqnarray}
\label{6.h5}
\lefteqn{\tau \frac{dh}{dt}=\pi (\tau \omega )^2\frac{\beta \tau }{rT(h)}\frac{2h}{(\tau \omega )^2}}\nonumber \\
& &{}\times \biggl\{ (1-\phi _\mu )\Lambda _1\left[ \frac{2h}{(\tau \omega )^2}\right] -\frac \beta r\Lambda _2\left[ \frac{2h}{(\tau
\omega )^2}\right] -\phi _g\biggr\},
\end{eqnarray}
where the functions
\begin{equation}
\label{6.h6}\Lambda _1[x]=\frac 2{\sqrt{1-x}(1+\sqrt{1-x})},
\end{equation}
\begin{equation}
\label{6.h7}\Lambda _2[x]=\frac{2x(1+2\sqrt{1-x})}{(1-x)^{3/2}(1+\sqrt{1-x}%
)^2}.
\end{equation}

In agreement with the results obtained in Sec.~\ref{sec:qa} from~(\ref
{6.h5}) we can see that the oscillatory instability occurs when $\phi _\mu +\phi
_g<1$ and is subcritical for $\phi _\mu +2\beta /r<1$. As follows from (\ref
{6.h5}) the dependence of the ``energy'' $h_{\text{st}}$ of steady state
oscillations on the physical parameters is specified by the expression
\begin{equation}
\label{6.h8}\phi _g=(1-\phi _\mu )\Lambda _1\left[ \frac{2h_{\text{st}}}{%
(\tau \omega )^2}\right] -\frac \beta r\Lambda _2\left[ \frac{2h_{\text{st}}%
}{(\tau \omega )^2}\right].
\end{equation}
In particular, for $\beta /r\ll
1-\phi _\mu \ll (\beta /r)^{3/2}$ and $\phi _g\ll [(1-\phi _\mu )(r/\beta
)^{1/3}]^{3/2}$ the value
\begin{equation}
\label{6.h9}h_{\text{st}}\approx \frac{(\tau \omega )^2}2\left[ 1-\frac \beta
{r(1-\phi _\mu )}\right].
\end{equation}
Besides, according to (\ref{6.h}%
) the maximum $a_{\text{max}}$ and the minimum $a_{\text{min}}$ attained by
the variable $a$ during oscillations are $a_{\text{max}}\approx a_{\text{max}%
}\approx (2h_{\text{st}})^{1/2}$. Whence for the given values of $\phi _g$
and $\phi _\mu $ we obtain
\begin{equation}
\label{ad.2}
\begin{array}{ll}
u_{\text{min}}\approx -\frac 12, & a_{
\text{min}}\approx \tau \omega , \\ u_{\text{max}}\approx \frac{2r}\beta
(1-\phi _\mu ), & a_{\text{max}}\approx \tau \omega .
\end{array}
\end{equation}
The period $T$ of these oscillations is practically determined by the time $%
\tau _s^{-}$ it takes for the system to pass through the region $u<0$.
Equation~(\ref{4.9}) enables us to estimate this time as $\tau _s^{-}\approx
\tau a_{\text{max}}/(\tau \omega )^2$ whence we get
\begin{equation}
\label{ad.3}T\sim \frac \tau {\tau \omega }.
\end{equation}
In the case under consideration the characteristic relative position of the
nullcline $a_0[u]$ and the limit cycle of the oscillations are demonstrated
in Fig.~\ref{F.7}.

In limits (\ref{6.100a}) and (\ref{6.100b}) the interface oscillations are
strongly--dissipative. The corresponding phase paths as a whole cannot be
described by level curves of any function similar to the energy ${\cal H}%
(u,a)$. In particular, in the region $u\lesssim 1$ these phase paths may be
roughly related to the unclosed level lines of the energy ${\cal H}(u,a)$.
In order to analyze the interface oscillations in this case we may make use
of the classic theory of relaxation oscillations \cite{AVK66}. First, we
consider in detail limit (\ref{6.100a}). Under these conditions, as it can
be shown directly from the system of equations~(\ref{4.8}), (\ref{4.9}), the
phase path going through a point that is not located in the region $|u+1|\ll
1$ or in a small neighborhood of the nullcline $a=a_0[u]$ forms practically
a horizontal line, at least, in the vicinity of this point. In other words,
at such a point the value $du/dt$ is large in comparison with $da/dt$ and
the quantity $\tau \omega $ may be treated as a small parameter. Therefore,
the interface dynamics should include fast motion governed by equation~(\ref
{4.8}) for fixed $a$ until the phase path reaches a small neighborhood of
the nullcline $a=a_0[u]$, where the left--hand side of equation~(\ref{4.8})
tends to zero, or a small neighborhood of the boundary $u=-1$, which
formally can be also regarded as the other branch of nullcline of equation~(%
\ref{4.8}). In the two latter regions the interface motion is slow.

The nullcline $a=a_0[u]$ (shown by curve 1 in Fig.~\ref{F.9}) looks like an
``N'' and the stationary point $(u=0,a=0)$ is unstable when it belongs to
the decreasing segment of the nullcline. So according to the classic theory
of relaxation oscillations \cite{AVK66} one could expect that the limit
cycle will be of the form presented by the dashed line in Fig.~\ref{F.9}.
Indeed, based on the system of equations (\ref{4.8}), (\ref{4.9}) we can
show that the increasing segments I, III of the nullcline $a=a_0[u]$ are
formally attractive for the phase path whereas the increasing one is
(segment II) is repulsive. The main characteristics of such a limit cycle
are directly determined by the form of the nullcline. In particular, the
minimum and maximum attained by the variable $a$ during oscillations are
approximately equal to $\tilde a_{\text{min}}$ and $\tilde a_{\text{max}}$,
i.e. the $a$--coordinates of the extremum points ($\tilde u_{\text{min}},%
\tilde a_{\text{min}}$), ($\tilde u_{\text{max}},\tilde a_{\text{max}}$) of
the nullcline. In the case under consideration the curve $a_0[u]$ is
specified by expression (\ref{ncsu}) and the last term on its right--hand
side is ignorable for $u\gg 1$. Whence we find that the quantities $\tilde a%
_{\text{min}},\tilde u_{\text{min}}$ can be estimated by expressions (\ref
{ad.2}) and for $\tilde a_{\text{max}}, \tilde u_{\text{max}}$ we obtain
\begin{equation}
\label{6.h10}|\tilde u_{\text{max}}|\lesssim 1 \quad 0< \tilde a_{\text{max}%
}\lesssim \frac{(1-\phi _\mu )\beta}r
\end{equation}
However, the right--hand side of equation (\ref{4.8}) possesses a certain
peculiarity in the region $|u+1|\ll 1$, because it tends to zero as $%
u\rightarrow -1$. Therefore, if the phase path goes into this region the
time scale hierarchy will be changed and, thus, the phase path will not be
able to follow the segment of the nullcline $a=a_0[u]$ going through this
region, and the maximum $a_{\text{max}}$ attained by the variable $a$ during
oscillations will be not determined by the extremum point ($\tilde u_{\text{%
max}},\tilde a_{\text{max}}$). This is the case in limit (\ref{6.100a})
because for such values of the physical parameters the nullcline $a=a_0[u]$
reaches a small neighborhood of the line $u=-1$ in the region $a>0$ and the
phase path has to go into the region $|u+1|\ll 1$ returning from large
values of $u$. Therefore, under the given conditions the limit cycle of the
interface oscillations will involve the conventional segments going along
the horizontal lines $a=a_{\text{max}}$, $a=\tilde a_{\text{min}}$, along
segment III of the nullcline $a=a_0[u]$ from $a_{\text{max}}$ to $\tilde a_{%
\text{min}}$ in addition to a certain anomalous segment located in the
region $|u+1|\ll 1$ which joins the two horizontal segments.

In order to complete the limit cycle construction we need to analyze the
system motion in the region $|u+1|\ll 1$ by solving directly the system of
equations (\ref{4.8}), (\ref{4.9}). Dividing equation~(\ref{4.8}) by
equation~(\ref{4.9}), setting $u=-1$ except for the terms containing the
cofactor $(1+u)$, and taking into account that in the given case $a\ll 1$ we
obtain the following equation describing the phase path in the region $%
|1+u|\ll 1$%
\begin{equation}
\label{6.hh1}\frac{du}{da}=(1+u)^3\left[ \frac r{\beta \phi _g}a+\frac 1{%
(1+u)^2}-\frac{1-\phi _\mu }{\phi _g}\right]
\end{equation}
subject to the formal ``initial'' condition
\begin{equation}
\label{6.hh2}u\rightarrow \infty \quad \text{as}\quad a\rightarrow \tilde a_{%
\text{min}}+0.
\end{equation}
Condition (\ref{6.hh2}) reflects the fact that the limit cycle segment
located in the region $|1+u|\ll 1$ originates from the phase path going into
this region practically along the horizontal line $a=\tilde a_{\text{min}}$.
To second order in $a$ the solution of equation (\ref{6.hh1}) meeting (\ref
{6.hh2}) is of the form
\begin{equation}
\label{6.hh3}\frac 1{(1+u)^2}\approx \frac r{\beta \phi _g}(a-\tilde a_{%
\text{min}})(-a-\tilde a_{\text{min}}).
\end{equation}
As it must $u\rightarrow \infty $ as $a\rightarrow \tilde a_{\text{min}}+0$,
the second point where the value $u$ tends to infinity is $a=-\tilde a_{%
\text{min}}$. Besides, according to (\ref{6.hh3}), values of $u$ such that $%
(1+u)\sim 1$ correspond to values of $a$ belonging to small neighborhoods of
the points $-\tilde a_{\text{min}}$ and $\tilde a_{\text{min}}$, i.e. $|a+%
\tilde a_{\text{min}}|\ll |\tilde a_{\text{min}}|$ or $|a-\tilde a_{\text{min%
}}|\ll |\tilde a_{\text{min}}|$ because in the given case the ratio $r\tilde
a_{\text{min}}^2/(\beta \phi _g)$ is a large parameter. Whence it follows
that the desired value of $a_{\text{max}}$ can be estimated as $a_{\text{max}%
}\approx -\tilde a_{\text{min}}$ i.e.
\begin{equation}
\label{6.hh4}a_{\text{max}}\approx \frac{(1-\phi _\mu )^2}4.
\end{equation}
It should be noted that $a_{\text{max}}\gg \tilde a_{\text{max}}$, so the
limit cycle goes over the extremum point ($\tilde u_{\text{max}},\tilde a_{%
\text{max}}$) at a remarkable distance. In addition, formula (\ref{6.hh3})
shows that the minimum $u_{\text{min}}$ attained by the variable $u$ during
oscillations is about $u_{\text{min}}\approx -1+[(\beta \phi _g)/(r\tilde a_{%
\text{min}}^2)]^{1/2}$, thus
\begin{equation}
\label{6.hh5}u_{\text{min}}\approx -1+\left[ \frac{4\beta \phi _g}{r(1-\phi
_\mu )^4}\right] ^{1/2}.
\end{equation}
The maximum $u_{\text{max}}$ attained during oscillations is determined by
the intersection point of the horizontal line $a=a_{\text{max}}$ and the
nullcline $a=a_0[u]$, whence we get
\begin{equation}
\label{6.hh6}u_{\text{max}}\approx \frac{(1+\sqrt{2})(1-\phi _\mu )r}{2\beta
}.
\end{equation}

Let us now estimate the characteristic time scales of the interface
oscillations. According to the classic theory of relaxation oscillations
\cite{AVK66} the fast motion along the horizontal lines $a=\tilde a_{\text{%
min}}$ and $a=a_{\text{max}}$ governed by equation~(\ref{4.8}) is
characterized by the time scale $\tau _f\sim \tau /a_{\text{max}}$. The time
during which the system moves slowly inside the region $|1+u|\ll 1$
practically in the $a$--direction from the point $\tilde a_{\text{min}}$ to $%
a_{\text{max}}$ is about $\tau ^{-}_s\sim 2a_{\text{max}}\tau /(\tau \omega
)^2\sim \tau (1-\phi _\mu )^2r/(\beta \phi _g)\gg \tau _f$. The motion along
segment III of the nullcline $a=a_0[u]$ is actually governed by equation~(%
\ref{4.9}) where we may roughly set $u\sim u_{\text{max}}$. In this way we
find that for this stage of the interface motion the characteristic time can
be estimated as $\tau ^{+}_s\sim \tau ^{-}_s/u_{\text{max}}\sim \tau (1-\phi
_\mu )/\phi _g\sim \tau $. It should be pointed out that in spite of $\tau
^{+}_s\ll \tau _f\sim \tau /a_{\text{max}}$ the motion along the nullcline $%
a_0[u]$ in the region $u\sim u_{\text{max}}$ is slow in comparison with the
fast motion along the horizontal lines $a=a_{\text{max}}$ or $a=a_{\text{min}%
}$ because in this region the time scale of the fast motion is $\tau _f/u^3_{%
\text{max}}$ rather than $\tau _f$. In the given case the period $T$ of the
system motion is determined, as before, by the value $\tau ^{-}_s$, thus
\begin{equation}
\label{ad.4}T\sim \tau \frac{r}{\beta}\frac{(1-\phi _{\tau})^2}{\phi _g}
\end{equation}

Let us now consider the interface oscillations in limit (\ref{6.100b}). In
addition we shall assume that $\phi _\mu \ll \phi _{\mu b}=1/6$ (Sec.~\ref
{sec:qa}), because for $\phi _\mu <\phi _{\mu b}$ the oscillation onset is
subcritical. In this case $\mu \ll 1$, the $k[u]$--dependence and the
nullcline $a_0[u]$ may be approximately specified by the expressions
\begin{equation}
\label{6.hh10}k[u]\approx k[0](1+u)
\end{equation}
and
\begin{equation}
\label{6.hh11}a_0[u]\approx -\frac u{1+u}+\phi _g\frac u{(1+u)^3}+\phi _\mu
u.
\end{equation}

For $\phi _g\sim 1$ the system of equations (\ref{4.8}), (\ref{4.9}) does
not contain actually a small parameter. Nevertheless, the minimum $\tilde a_{%
\text{min}}$ attained by the nullcline $a_0[u]$ in the region $u>0 $ is
located in the immediate vicinity of the boundary $a=-1$, i.e. $\tilde a_{%
\text{min}}$ $\approx -1$ and matches the value $\tilde u_{\text{min}}\sim
1/(\phi _\mu )^{1/2}\gg 1$. Besides, the increasing segment of the
nullcline $a_0[u]$ (for $u>0$) corresponds to large values of the variable
$u$.
Therefore, in the given case the variable $u$ attains large values
during interface oscillations and the limit cycle comes close to the line $%
a=-1$. Since the right--hand side of equation~(\ref{4.8}) depends on $u$ and
$a$ as $(1+u)^4/(1+a)$ this behavior of the limit cycle causes the interface
oscillations to be relaxation and the limit cycle contains segments of the
fast motion that approximately are parallel to the $u$--axis, a segment
following the increasing part of the nullcline $a=a_0[u]$ in the region $%
u\gg 1$, and a segment located in the region $|u+1|\lesssim 1$. The latter
one, however, also approximately follows the corresponding increasing part
of the nullcline $a=a_0[u]$ as results from numerical analysis
although the bifurcation is subcritical for the given physical
parameters. This characteristics explains the fact that the region of the
parameter $\phi _g$ where the interface motion is linearly--stable and the
interface oscillations with a finite amplitude can occur is sufficiently
narrow. In particular, as formally $\phi _\mu \rightarrow 0$ this region is
determined by the inequality $1<\phi _g<\phi _{gc}\approx 1.05 $.

Inside the instability region far from the neutral curve, i.e. for $\phi
_g\ll 1$, the limit cycle again will reach the region $|u+1|\ll 1$, thus the
corresponding segment of the limit cycle will not follows the increasing
part of the nullcline $a=a_0[u]$ and take a form similar to that matching
limit (\ref{6.100a}). This conclusion has been obtained by solving
numerically the system of equation (\ref{4.8}), (\ref{4.9}) in the region $%
|u+1|\ll 1$ for $\phi _g\ll 1$ where it can be rewritten in the
parameterless form
\begin{mathletters}
\begin{eqnarray}
\label{6.hh15a}
\frac{\partial u_p}{\partial t}&=&\frac{u_p^4}a_p\left[ a_p-\frac
1u_p(1-\frac 1{u_p^2})\right] , \\
\label{6.hh15b}\frac{\partial a_p}{\partial t}&=&1,
\end{eqnarray}
\end{mathletters}
where $u_p=(u+1)/\phi _g^{1/2}$, $a_p=(a+1)\phi _g^{1/2}$, and the time $t$
is measured in units of $\tau /\phi _g^{3/2}$. In
the given region the phase path substantially deviates from the nullcline
and goes over its extremum point ($\tilde u_{\text{max}},\tilde a_{\text{max}%
}$) at a certain distance. So also in the case
under consideration the limit cycle of the interface oscillations and one
formally predicted by the classic theory of relaxation oscillations \cite
{AVK66} differ significantly in form. The period of such oscillations can be
roughly estimated as
\begin{equation}
\label{add.10}T\sim \frac{\tau}{\phi ^{3/2}_g}.
\end{equation}

Concluding the present section we note that, first, in all the cases
considered hear the interface oscillations may be regarded as spikewise. The
matter is that the interface dynamics described in terms of the variables $a$
and $u$, i.e. by the system of equations (\ref{4.8}), (\ref{4.9}) is
characterized by self--acceleration in the region $u\ll 1$, which is
reflected in the existence of the cofactor $(1+u)^3$ in equation (\ref{4.8}%
). So the interface dynamics is characterized, at least, by two time scales,
$\tau ^{-},\tau ^{+}$, corresponding to the motion in the regions $%
|u|\lesssim 1$ and $u\gg 1$, respectively, with $\tau ^{-}\gg \tau ^{+}$.
Therefore, the time course of these oscillations contains spikes of duration
$\tau ^{+}$ formed during motion in the region $u\gg 1$. Second, from the
standpoint of the crystal growth theory it is important to analyze the main
characteristics of impurity distribution in the growing crystal. So we also
estimate the spatial period $H$ of the impurity distribution $C_{\text{sol}%
}(y)$ occurring during strongly--dissipative interface oscillations. By
definition

\begin{equation}
\label{6.hh16}T_y=V\int_0^Tdt(1+u).
\end{equation}
For example, in limit (\ref{6.100a}) the main contribution to the value of
integral (\ref{6.hh16}) is due to the system motion along segment III of the
nullcline. Then substituting the corresponding estimates into (\ref{6.hh16})
we get $H\sim Vu_{\text{max}}\tau _s^{+}$, so
\begin{equation}
\label{add.11}H\sim \frac D{Vk(V)}\frac{r(1-\phi _\mu )^2}{\beta \phi _g}
\end{equation}
In case (\ref{6.100b}) for $\phi _g\lesssim 1$ and $\phi _\mu \ll 1$ the
spatial period is about
\begin{equation}
\label{add.12}H\sim \frac D{Vk(V)}
\end{equation}
The comparison of (\ref{add.11}) and (\ref{add.12}) shows that expression~(%
\ref{add.12}) can be treated as a rough estimate of the spatial period of
the impurity distribution occurring during strongly--dissipative interface
oscillations. Keeping in mind the layer--by--layer crystal growth we set $%
D\sim 5\cdot 10^{-5}$ cm$^2$/sec, $V\sim 10^{-2}$ cm/sec and $k(V)\sim 0.1$
for such values of $V$. Then from (\ref{add.12}) we get $H\sim 500$ $\mu $m,
which coincides in order with the characteristic length of impurity
microinhomogeneities observed in real crystals \cite{M91}.

In addition, we consider some characteristic features of the $C_{\text{sol}%
}(y)$ dependence. Expressions (\ref{4.6}), (\ref{4.7}) lead us to the
following relationship
\begin{eqnarray}
\label{6.20}
\frac{C_{\text{sol}}}{C_{\infty}} & = & 1+\frac{k[u]}{k[0]}(\mu +1)
(a-a_0[u])+(\tau \omega)^2\frac{u}{(1+u)^2}\nonumber \\
& &\mbox{} = \frac{k[u]}{k[0]} [1+(1+ \mu )a - \mu u ].
\end{eqnarray}
During the system motion practically along segment III of the nullcline $%
a_0[u]$ the concentration $C_{\text{sol}} $ is actually equal to $C_{\infty}$
because the right hand side of (\ref{6.20}) differs from one by the value $%
(\tau \omega)^2u/(1+u)^2 \ll 1$ for $u \gtrsim 1$. As seen from expression (%
\ref{6.20}) in cases where $\mu \ll 1$ the impurity concentration in the
solid near the interface attains its minimum $C_{\text{sol}}^{\text{min}}$
and maximum $C_{\text{sol}}^{\text{max}}$ when in the $ua$--plane the system
reaches the points with the coordinates $a\sim a_{\text{min}},u\sim u_{\text{%
min}}$ and $a\sim a_{\text{max}},u\sim u_{\text{max}}$, respectively. So the
minimum and the maximum of the impurity distribution $C_{\text{sol}}(y)$ can
be estimated as
\begin{eqnarray*}
C_{\text{sol}}^{\text{min}}&\sim & C_{\infty}\frac{k[u_{\text{min}}]}{k[0]}(1+a_{\text{min}}), \\
C_{\text{sol}}^{\text{max}}&\sim & C_{\infty}\frac{k[u_{\text{max}}]}{k[0]}(1+a_{\text{max}}).
\end{eqnarray*}

In order to analyze the feasibility of the quasistationary approximation for
the interface relaxation oscillations we have also solved the system of
equations (\ref{4.8})-(\ref{4.9}) as well as the full one-dimensional model (%
\ref{2.1})-(\ref{2.7}) numerically. We used the following values of the
parameters: $k_0=0.0125,\,k_\infty =0.8,\,\beta ^{*}V=0.125,\,Mk=10$, and $%
\mu =0.2$, which correspond to the subcritical bifurcation and are not too
far from the critical point. The obtained results are presented in Fig.~\ref
{F.11}--Fig.~\ref{F.13}.

It should be expected that the quasistationary approximation may be violated
when the interface velocity becomes small enough: $(1+u)\ll 1$.
Nevertheless, as seen from these figures, the shapes of the limit cycle
constructed by different methods qualitatively coincides with each other.
Therefore, first, based on the quasistationary approximation one can
describe the main characteristics of oscillatory zoning. Second, the
estimates for the amplitude and spacing of the growing superlattice can be
used as a first approximation for the real processes.

\section*{ACKNOWLEDGMENTS}

The authors thank Professor V.V. Voronkov for interest and useful
discussions. The research described in the present paper was made possible
in part by Grants No. U1I000, U1I200 from the International Science
Foundation.

\begin{figure}

\begin{figure}
\caption{Geometry of the stability region in the $C_\infty V$--plane
depending on the surface tension. (The interface is stable at points below
the solid curve. Figures (a), (b), and (c) correspond to the conditions
$d\ll d_1$, $d\sim d_1$, and $d\gg d_1$, respectively. Curves I and II are
the boundaries of the marginal and oscillatory instabilities.)}
\label{F.3}
\end{figure}
\caption{Relative position of the absolute stability boundary (curve~1),
the neutral curve (curve~2), and the curve of the transition
between the supercritical and subcritical bifurcation (curve~3).}
\label{F.2}
\end{figure}

\begin{figure}
\caption{The dimensionless front velocity $u$ (a) and the dimensionless
interface position $a$ (b) vs. time (in units $D/V^2$) for the
supercritical bifurcation.}
\label{F.4}
\end{figure}

\begin{figure}
\caption{The interface oscillations and the nullcline for the supercritical
bifurcation (the thick solid and dotted lines are the limit cycle obtained
within the framework of the full diffusion problem and in the
quasistationary approximation, respectively).}
\label{F.5}
\end{figure}

\begin{figure}
\caption{Distribution of the impurity concentration $C_{\text{sol}}$ in the
growing crystal for the supercritical bifurcation (the concentration
$C_{\text{sol}}$ and coordinate $y$ are measured in units $C_{\infty}$ and
$D/V$).}
\label{F.6}
\end{figure}

\begin{figure}
\caption{Schemetic illustration of the phase path geometry in the $ua$--plane
in limit ({\protect \ref{6.100}}) (the thick line represents the nullcline
$a=a_0[u]$ and the dashed line separates the regions of closed and unclosed
phase paths).}
\label{F.7}
\end{figure}

\begin{figure}
\caption{Characteristic geometry of the limit cycle in the $ua$--plane for
the subcritical bifurcation (the solid line is the nullcline $a_0[u]$,
the thick and dashed lines correspond to the limit cycle constracted
analitically and predicted by the classic theory of relaxation oscillations,
respectively).}
\label{F.9}
\end{figure}

\begin{figure}
\caption{The interface oscillations and the nullcline for the subcritical
bifurcation (the thick and dotted lines are the limit cycle obtained within
the framework of the full diffusion problem and in the quasistationary
approximation, respectively).}
\label{F.11}
\end{figure}

\begin{figure}
\caption{The dimensionless front velocity $u$ (a) and the dimensionless
interface position $a$ (b) vs. time (in units $D/V^2$) for the
subcritical bifurcation.}
\label{F.12}
\end{figure}

\begin{figure}
\caption{Distribution of the impurity concentration $C_{\text{sol}}$ in the
growing crystal for the subcritical bifurcation (the concentration
$C_{\text{sol}}$ and coordinate $y$ are measured in units $C_{\infty}$ and
$D/V$).}
\label{F.13}
\end{figure}

\end{document}